\documentclass[12pt]{article}

\usepackage{graphicx, xspace, floatflt}
\usepackage{multicol}
\usepackage{natbib}

\def\cm3{{cm$^{-3}$}}
\def\longrefsmall#1 {\par{\hangindent=10pt \hangafter=1 #1 \par}}

\def\la{\ifmmode\stackrel{<}{_{\sim}}\else$\stackrel{<}{_{\sim}}$\fi} 
\def\ga{\ifmmode\stackrel{>}{_{\sim}}\else$\stackrel{>}{_{\sim}}$\fi}

\begin{document}

\pagenumbering{arabic}
\begin{center}
\vspace*{0.1in}
{\bf New Discoveries in Planetary Systems and Star Formation
    through Advances in Laboratory Astrophysics}
\vskip 0.2in
Submitted by the 
\vskip 0.2in
American Astronomical Society Working Group on Laboratory Astrophysics\\
http://www.aas.org/labastro/
\vskip 0.2in
Nancy Brickhouse$^*$ - Harvard-Smithsonian Center for
    Astrophysics \\
nbrickhouse@cfa.harvard.edu, 617-495-7438\\
\vskip 0.2in
John Cowan - University of Oklahoma\\
cowan@nhn.ou.edu, 405-325-3961\\
\vskip 0.2in
Paul Drake - University of Michigan \\
rpdrake@umich.edu, 734-763-4072
\vskip 0.2in
Steven Federman - University of Toledo\\
steven.federman@utoledo.edu, 419-530-2652
\vskip 0.2in
Gary Ferland - University of Kentucky\\
gary@pa.uky.edu, 859-257-8795\\
\vskip 0.2in
Adam Frank - University of Rochester\\
afrank@pas.rochester.edu, 585-272-1717
\vskip 0.2in
Eric Herbst - Ohio State University\\
herbst@mps.ohio-state.edu, 614-292-6951
\vskip 0.2in
Keith Olive - University of Minnesota\\
olive@physics.umn.edu, 612-624-7375\\
\vskip 0.2in
Farid Salama - NASA/Ames Research Center\\
Farid.Salama@nasa.gov, 650-604-3384\\
\vskip 0.2in
Daniel Wolf Savin - Columbia University\\
savin@astro.columbia.edu, 212-854-4124\\
\vskip 0.2in
Lucy Ziurys - University of Arizona\\
lziurys@as.arizona.edu, 520-621-6525
\end{center} 
\vskip 0.2in

$^*$Editor

\clearpage


\begin{center}
{\bf 1. Introduction}
\end{center}

\vspace{-0.15in}
As the panel on Planetary Systems and Star Formation (PSF) is fully
aware, the next decade will see major advances in our understanding of
these areas of research. To quote from their charge, these advances
will occur in studies of ``solar system bodies (other than the Sun)
and extrasolar planets, debris disks, exobiology, the formation of
individual stars, protostellar and protoplanetary disks, molecular
clouds and the cold ISM, dust, and astrochemistry.''

Central to the progress in these areas are the corresponding advances
in laboratory astrophysics which are required for fully realizing the
PSF scientific opportunities in the decade 2010-2020.  Laboratory
astrophysics comprises both theoretical and experimental studies of
the underlying physics and chemistry which produce the observed spectra and describe
the astrophysical processes.  We discuss four areas of
laboratory astrophysics relevant to the PSF panel: atomic,
molecular, solid matter, and plasma physics.

Section 2  describes some of the new
opportunities and compelling themes which will
be enabled by advances in laboratory astrophysics.  Section 3 
provides the scientific context for these opportunities.  Section 4
discusses some experimental and theoretical advances in
laboratory astrophysics required to realize the PSF scientific
opportunities of the next decade.  As requested in the Call for White
Papers, we present in Section 5  four central questions and one area with
unusual discovery potential.  We give a short postlude in
Section 6.

\vspace{-0.1in}
\begin{center}
{\bf 2. New scientific opportunities and compelling scientific themes}
\end{center}

\vspace{-0.15in}
In the upcoming decade, we will make progress in understanding the
evolution from clouds of gas and dust to systems of stars and
planets. Studies of protostellar and protoplanetary disks and
debris disks will inspire new ideas to explain how planets
form and migrate within young systems. We will not only continue to
discover new and interesting planets but we will also begin to
characterize and classify the properties of these planets and their
atmospheres and cores. The discoveries of earth-like planets and other
planets in habitable zones will challenge us to address the origin of
life.   We expect to solve long-standing mysteries,
such as the nature of the diffuse interstellar bands (DIBs), believed
to be the manifestation of an unidentified reservoir of organic
material.  Discovery of the carriers of these bands promises to
revolutionize our understanding of organic chemistry in the
interstellar medium (ISM).

Perhaps the most exciting new opportunities will come in the area of
the origins of life. The cosmic pathway to life begins in interstellar
gas clouds where atomic carbon is “fixed” into molecules, thereby
initiating the synthesis of the complex organic species that are
eventually sequestered on planets.  These reactions initiate not only
the formation of organic molecules in the cosmos, but also provide
some of the first threads knitting atoms and molecules into solid
material. Such processes are critical for the eventual formation of
planets and may determine a major component of the organic chemistry
that is present on their young surfaces. To place the new
worlds we discover in proper context requires that we
understand the chain of events that dictates their
potential chemical inventories.

\vspace{-0.1in}
\begin{center}
{\bf 3. Scientific context}
\end{center}

\vspace{-0.15in}
With a multitude of current and planned space-based and ground-based
telescopes, including {\it Spitzer}, {\it Hubble}, {\it Chandra}, {\it
XMM-Newton}, {\it Suzaku}, {\it Herschel}, {\it SOFIA}, {\it Kepler},
Keck, Magellan, the MMT, Hobby-Eberly, SALT, LBT, the Gemini
telescopes, and eventually ALMA and {\it JWST}, the upcoming decade
promises numerous opportunities for progress in the areas of planet
and star formation. Studies from the radio to X-rays contribute to our
overall understanding of the physical and chemical processes.

Understanding the process of accretion is key to astrophysics from the
formation of the lowest mass stars to the growth of super-massive
black holes. Star formation studies over the past decade have begun to
recognize the importance of the interaction between a young star and
its environment. X-ray and ultraviolet (UV) radiation produced by a
low mass star can have profound effects on the ionization and heating
of gas and dust in the disk (Glassgold et al. 2004; Jonkheid et
al. 2004). The star's magnetic fields can extend to the disk and help
to channel the accreting material. Flares are ubiquitous in young
low-mass stars (Feigelson et al. 2007) and thus the inner disk
experiences highly time-variable phenomena. In addition to disk-driven
winds, magnetic accretion can generate jets and winds.

The explosion of exoplanet discoveries over the past decade suggests
that the next decade will revolutionize our understanding of the
formation of planets, their atmospheres and cores, and the rich
diversity of planetary systems. The transit method of exoplanet
searching, at the core of {\it Kepler's} goal to discover other
Earths, requires accurate and finely-tuned atmosphere models. The
first direct studies of light emitted by an exoplanet itself (Deming
et al. 2005; Charbonneau et al. 2005) estimate temperatures from
simple models fit to broad-band infrared (IR) observations; water has
now been observed spectroscopically on an exoplanet (Grillmair et
al. 2008).  In the next decade observations will continue to drive the
sophistication of planet atmosphere models (e.g. winds from close-in
Jupiters; Murray-Clay et al. 2008). Gas giants dominate our planetary
system by mass and may be ubiquitous in stellar systems. A favored
model for their formation requires a metallic core; however, it is not
clear whether the heavy core survives, an issue which can be addressed
by experimental studies of high density plasmas.

The evolution of gas and mineral dust in protoplanetary disks traces
the formation process of planets and other objects. Comparisons of
exoplanet observations with studies of objects in our own solar system
will help to determine the common evolutionary histories (e.g. Bergin
et al. 2007). Discoveries of smaller planets and multi-planet systems
will shed light on the timescales for formation, migration,
destruction, and the relative roles of dust and gas. The growth of
dust grains in the presence of stellar radiation depends on the
gas-to-dust ratio and the dust temperature which, in turn, control the
chemistry. Gas and dust are coupled at the high densities found in
the mid-plane of the disk, suggesting that some molecules are frozen
onto grains. Disk dynamics can alter the properties of gas and dust
through mixing. 

The discovery of over 140 different chemical molecules in interstellar
gas, with the vast majority organic molecules, and the realization
that obscuring dust pervades vast regions of the ISM have revealed the
complexity of interstellar chemistry. The first stars are thought to
have formed from primordial clouds where H$_2$ and HD controlled their
cooling and collapse (Bromm et al. 2002).  Subsequent
stars and planetary systems have formed out of the most complex
molecular environments deep inside cold gas and dust clouds, often
obscured by hundreds of visual magnitudes of extinction. All the main
functional groups known to organic chemists have now been observed in
interstellar molecules, suggesting that interstellar chemistry
contains the organic complexity seen on Earth. The next decade may
establish the gas phase and grain chemistries of interstellar clouds
that can lead to the origin of life.

\clearpage
\begin{center}
{\bf 4. Required advances in Laboratory Astrophysics}
\end{center}

\vspace{-0.15in}
Spectroscopic studies are key to identifying the constituents of
various environments, diagnosing the physical and chemical conditions,
and testing models of the physical processes and chemical reactions. A
strong program in laboratory astrophysics allows us to make steady
progress in compiling databases of the spectral properties of neutral
atoms and ions, molecules, and minerals, and to react quickly to
address surprises from new observations.  While laboratory
astrophysics has contributed to many successes in
astrophysics,\footnote{http://www.physics.unlv.edu/labastro/} in many
areas of ongoing research in PSF fields, we know that the linelists
are not complete enough and that the rate coefficients are not
accurate enough to take full advantage of the wealth of new data
expected over the next decade.  Moreover, models of astrophysical
plasma processes should be tested experimentally to ensure that the
assumptions are valid.

\vspace{-0.1in}
\begin{center}
{\bf 4.1. Atomic Physics}
\end{center}

\vspace{-0.15in}
Understanding the star-disk interaction
in protostellar and young stellar systems requires
accurate knowledge of the atomic physics and fundamental spectroscopy
across many wavebands. X-ray emission from an accretion shock has
been identified by X-ray spectral line ratios indicative of
high electron density (Kastner et al. 2002). Unprecedented accuracy
has now been demonstrated for atomic theory and experiment (Chen et
al. 2006), but much more work is needed to generate a comprehensive
set of reliable diagnostics for temperature and density. Stellar X-ray
and UV radiation can excite specific lines in the disk, such  Fe
K$\alpha$  (Imanishi et al. 2001) and [Ne II] 12.8 $\mu$  (Pascucci et al. 2007).  The processes involved in
understanding such emission include photoionization,
radiative and dielectronic recombination,
collisional excitation, X-ray fluorescence, Auger ionization, and
charge exchange. Accurate line lists, oscillator strengths, atomic
transition probabilities, and collisional rate coefficients are needed
to develop new diagnostics from the IR to the X-ray.

Planet searches require accurate stellar atmosphere
models for cross correlation with the observed spectra
(e.g. Konacki et al. 2004). Characterization of planetary atmospheres
places new demands on atmosphere models for accurate and complete
atomic line opacities. Ultimately the accuracy of the derived parameters
depends on the accuracy of the template spectra; where lines are
missing in the template, the observed spectrum cannot be
used.

\vspace{-0.1in}
\begin{center}
{\bf 4.2. Molecular Physics}
\end{center}

\vspace{-0.15in}

High-resolution laboratory spectroscopy is absolutely essential in
establishing the identity and abundances of molecules observed in
astronomical data. For molecules already identified in astronomical
sources, transitions including higher energy levels and new isotopic
species are needed as we move into new spectral regions and higher
resolving powers. Furthermore, the spectra of as yet undiscovered
species that promise to serve as important new probes of astronomical
sources need to be identified. For example, the recent astronomical discovery of
the negative ion C$_6$H$^-$ (McCarthy et al. 2006) was driven by
laboratory work. We anticipate that additional discoveries of anions in the
coming decade will challenge the existing view of interstellar chemistry.

New line diagnostics of disk heating, cooling, and ionization are
expected from continuing laboratory work on abundant molecules. For
example, submillimeter (sub-mm) lines of CO from high rotational levels have
recently been observed in the outer disk of a young star, requiring
additional heating such as by X-rays (Qi et al. 2006); 
numerous UV lines of H$_2$ were observed in the same system, pumped by
X-ray or UV radiation (Herczeg et al. 2002).

The ``unidentified line'' problem needs to be addressed by systematic
measurements of a variety of molecules. The DIBs were first observed
in 1919, but despite many decades of intense efforts by laboratory
spectroscopists and astronomers, the molecular carriers of these bands
remain a mystery. Mid-IR spectra of individual objects such as H II
regions, reflection nebulae, and planetary nebulae as well as the
general ISM are dominated by a set of emission features due to large
aromatic molecules (PAH bands).  Studies of the IR spectroscopy of
such molecules and their characteristic features which depend on
molecular structure and charge state are of key importance for our
understanding of this ubiquitous molecular component of the
ISM. Millimeter and sub-mm spectra contain hundreds of unidentified
gas-phase lines.  Intense laboratory work continues to be a vital
component in unlocking the mystery of the DIBs, the PAH emission
bands, and other observed, but unidentified, spectral features (Salama
1999). Missing molecular lines in the coolest stellar atmosphere
models and in the atmosphere models being developed for planets create
an acute problem for interpreting and exploiting astronomical
spectra. Discovery of biological molecules from planetary atmospheres
may occur in the next decade (see Robinson et al. 2008). 

A vast number of molecules which are difficult to detect by standard
means may exist in the ISM and stellar systems. The identification
of most interstellar/circumstellar species is done through
observations of emission spectra in the radio and millimeter,
with an observational bias in favor of molecules with large
dipole moments.  A big challenge for observation is to find ways to
detect molecules with small or zero  dipole moments. Laboratory
spectroscopy in other wavebands would be useful.

Understanding line profiles will also provide new tools for PSF
studies.  For example, pressure broadening of the strong resonance
absorption lines of Na and K, observed in brown dwarfs (Burrows et
al. 2001) and predicted in extra-solar gas giants, occurs through
collisions with neutrals, including molecular hydrogen (Seager \&
Sasselov 2000). Quantum collision calculations, benchmarked by
laboratory measurements, can provide accurate line profiles which can
serve as diagnostics of the temperature and pressure (e.g. Zhu et
al. 2005).

Interpreting spectral observations requires the use of sophisticated
astrochemical models, which include the ionization and dissociation
mechanisms (by radiation and cosmic rays) and the time-dependent
history of the clouds. To understand the chemical composition of these
environments and to direct future molecular searches in the framework
of new astronomical observatories, it is important to untangle the
detailed chemical reactions and processes leading to the formation of
new molecules in extraterrestrial environments.  Breakthroughs in our
understanding of the molecular universe are limited by uncertainties
in the underlying chemical data in these models (e.g. Glover et
al. 2006).  Laboratory spectroscopy is crucial for understanding
interstellar molecule formation in the gas phase and on grains.

Of particular importance are data for reactions of neutral atomic
carbon atoms with molecular ions which are critical in initiating
interstellar organic chemistry. Theory is limited to classical
methods, as fully quantum mechanical reactions for systems with four
or more atoms are beyond computational capabilities now and for the
foreseeable future.  Existing laboratory experiments have produced
ambiguous results owing to the extraordinary challenge of generating
and characterizing atomic carbon beams with no internal energy.  The
spectroscopic study of such molecules, many of which are complex
structures that cannot be produced in large abundance in the
laboratory, requires the development and application of
state-of-the-art ultra-sensitive spectroscopic instruments. Tracking
the cosmo-chemical pathway towards life requires cross-disciplinary
work at the intersection of astronomy, chemistry, biology, and
physics.

\vspace{-0.1in}
\begin{center}
{\bf 4.3. Solid Material}
\end{center}

\vspace{-0.15in}
Spectroscopy provides
unique information on stellar outflows and protoplanetary disks,
accreting envelopes, the surrounding medium, the radiation fields, and
planetary atmospheres. To be useful,
however, this tool requires an ability first to identify the lines of
gas phase molecules and solid state band features of grains, and then
to explain their appearance using reliable physical excitation
constants.  Laboratory measurements supplemented by theoretical
modeling provide this essential information.

An example of the problem of identification is the emission feature
near 21 $\mu$m, originally discovered in four objects with large far
IR excesses due to a circumstellar dust envelope surrounding a
carbon-rich central star (Kwok et al. 1989). A similar feature, but
with varying wavelength centroid, has
been observed with Spitzer in the Carina nebula, the supernova
remnant Cas A, and some H II regions and is likely associated with
dust grains composed of FeO and a mix of silicates (e.g. Rho et
al. 2008). 

Laboratory measurements of molecules and grains at different
temperatures and under various radiation environments are needed to
inform the search for specific elements. For example, oxygen is the third most abundant
element in the universe, and early models of gas clouds predicted that
O~I, O$_2$, and H$_2$O  should be relatively
abundant.  But observations over the past decade by SWAS, Odin, ISO,
and other instruments found that H$_2$O is about 100 times less
abundant than expected in star forming regions; O$_2$
is similarly scarce. Oxygen plays a crucial role in the
chemistry of star forming regions: pairing with carbon to make the
crucial species CO for cooling the clouds (and thus facilitating cloud
collapse), acting as a key ingredient in the chemical pathways that
make most complex organic species, and by joining with silicon or
hydrogen on the dust grains to form complex molecules. Recent calculations suggest that the missing oxygen may
be in the form of water ice. 

Studies of silicate and carbonaceous
dust precursor molecules and grains are required as are studies of the
interaction between dust and its environment, including radiation and
gas (Tielens 2005).  Studies of dust and ice provide a clear connection between
astronomy within and beyond the solar system.  Astronomical
observations and supporting laboratory experiments over a wavelength
range that extends from the X-ray domain to the UV, IR, and
sub-mm are of paramount importance for studies of the molecular
and dusty universe.  Observations at IR and sub-mm wavelengths
penetrate the dusty regions and probe the processes occurring deep
within them.  These wavelengths provide detailed profiles of molecular
transitions associated with dust.

At longer wavelengths, the continuum dust opacity is uncertain by an
order of magnitude.  IR spectral features of interstellar dust grains
are used to determine their specific mineral composition, hence their
opacities, which determine inferred grain temperatures and the masses
of dusty objects, including the ISM of entire galaxies.  Emission
bands from warm astronomical environments such as circumstellar
regions, planetary nebulae, and star-forming clouds lead to the
determination of the composition and physical conditions in regions
where stars and planets form.  The laboratory data essential for
investigations of dust include measurements of the optical properties
of candidate grain materials (including carbonaceous and silicate
materials, as well as metallic carbides, sulfides, and oxides) as a
function of temperature.

For abundant materials (e.g., forms of carbon such as PAHs), the
measurements should range from gas-phase molecules to nanoparticles to
bulk materials. The IR spectral region is critical for the
identification of grain composition, but results are also required for
the UV.  Previous
studies in the UV have focused on the only identified spectral feature
(at 2200 \AA), but COS on {\it Hubble} is expected to find UV spectral
signatures for many materials, including specific individual aromatics.

\vspace{-0.1in}
\begin{center}
{\bf 4.4. Plasma Physics}
\end{center}

\vspace{-0.15in}
Accretion from a disk orbiting a protostar
requires efficient transport of angular momentum through the disk, but
how this happens, and the extent to which it is dominated by nonlinear
hydrodynamic turbulence or magnetohydrodynamic (MHD) turbulence via
magnetorotational instability (MRI) (Balbus \& Hawley 1998), has been
much debated. Recent laboratory experiments show that pure
hydrodynamic turbulence probably cannot explain  the observed
fast accretion (Ji et al. 2006). Laboratory jets are also providing
tests of models, including studies on radiative cooling and
collimation (Lebedev et al. 2002), instabilities (Ciardi et
al. 2009), and dynamics and mixing (Foster et al. 2005).

How magnetic field changes its topology and releases its energy is an
outstanding problem in plasma astrophysics (Biskamp
2000). Reconnection has been proposed as a mechanism for releasing the
energy built up as a result of differential rotation between the star
and the disk (van Ballegooijen 1994). Only recently have reconnection
models been tested experimentally.  For example, the well-known
Sweet-Parker model was tested quantitatively in the laboratory (Ji et
al. 1998) 40 years after its birth, and the conjectured Hall effect
has also now been successfully verified in the laboratory (Ren et
al. 2005). 

Knowledge of planetary interiors depends on models that incorporate
the properties of the interior matter. As the interior pressure
approaches and exceeds 1 Mbar (0.1 TPa, 10$^{12}$ dynes/cm$^2$), the interior
matter ionizes and becomes a high-energy-density plasma. There are
important transitions involving molecular dissociation, ionization,
Fermi degeneracy, and other effects under conditions where standard
theoretical treatments for cold, condensed matter, or hot, ionized
plasma are invalid. This is precisely the region of interest for
planetary formation, structure and interior dynamics. Understanding
the compressibility of the material is essential to develop structural
models consistent with data. Understanding conductivity is essential
to understand planetary magnetospheres. The existence of a metallic
state of hydrogen has been confirmed in the laboratory (Weir et
al. 1996), but the full insulator-to-conductor transition has yet to
be explored. Understanding phase transitions and material mixing is
essential to know how gravitational potential energy contributes to
the energy balance. The anomalous radiation from nearby gas giant
planets may be due to the separation, condensation, and raining
downward of helium that is initially mixed with hydrogen (Guillot
1999). Alternatively, perhaps so-called pycnonuclear reactions
(Ichimaru \& Kitamura 1999) may release energy in dense, metallic
hydrogen and deuterium.

In the next decade, further results are expected from the
laboratory in the areas of MHD, jets and magnetic reconnection.
Experimental facilities that can create and study the states of matter
present in planetary interiors exist or are under construction both in
the US and worldwide.   What is required is to exploit and to extend
the diagnostic capabilities that can quantify the detailed properties
of matter in planetary interiors, including not only density and
pressure but also conductivity, viscosity, and other
characteristics.


\vspace{-0.1in}
\begin{center}
{\bf 5. Four central questions and one area with unusual discovery potential}
\end{center}


\vspace{-0.15in}
\begin{itemize}
\vspace{-0.15in}
\item How does a young star interact with its environment?

\vspace{-0.15in}
\item What are the characteristics of planetary systems and their
atmospheres and interiors?

\vspace{-0.15in}
\item How do the gas and dust in disks around stars evolve and how are
they affected by disk dynamics, stellar radiation, and grain growth?

\vspace{-0.15in}
\item What are the abundances, distribution and formation mechanisms of the chemical compounds
  found in the ISM?

\vspace{-0.15in}
\item[$\star$] What is the cosmo-chemical pathway from atoms in interstellar space
to life?
\end{itemize}


\vspace{-0.3in}
\begin{center}
{\bf 6. Postlude}
\end{center}

\vspace{-0.15in}
Laboratory astrophysics and complementary theoretical calculations are
part of the foundation for our understanding of PSF and will remain so
for many decades to come. From the level of scientific conception to
that of scientific return, our understanding of the underlying
processes allows us to address fundamental questions in these
areas. Laboratory astrophysics is constantly necessary to maximize the
scientific return from astronomical observations.

\vspace{-0.1in}
\begin{center}
{\bf References}
\end{center}

\noindent{\small
\begin{minipage}[t]{3.1in}

\vspace{-0.2in}

\noindent\hangindent=15pt\hangafter=1 Balbus, S., \& Hawley, J. 1998, RMP, 70, 1

\noindent\hangindent=15pt\hangafter=1 Bergin, E. et al. 2007, in Protostars and Planets V, 751

\noindent\hangindent=15pt\hangafter=1 Biskamp, D. 2000, Magnetic Reconnection in Plasmas, Cambridge U Press

\noindent\hangindent=15pt\hangafter=1 Bromm, V. et al. 2002, ApJ, 564, 23

\noindent\hangindent=15pt\hangafter=1 Burrows, A. et al. 2001, RMP, 73, 719

\noindent\hangindent=15pt\hangafter=1 Charbonneau, D. et al. 2005, ApJ, 626, 523

\noindent\hangindent=15pt\hangafter=1 Chen, G.-X. et al. 2006, PRA, 74, 42749

\noindent\hangindent=15pt\hangafter=1 Ciardi, A. et al., ApJ, 691, L147

\noindent\hangindent=15pt\hangafter=1 Deming, D. et al. 2005, Nature, 434, 740

\noindent\hangindent=15pt\hangafter=1 Feigelson, E. et al. 2007, in Protostars and Planets V, 313

\noindent\hangindent=15pt\hangafter=1 Foster, J. M. et al. 2005, ApJ, 634, L77

\noindent\hangindent=15pt\hangafter=1 Glassgold, A. E.  et al. 2004, ApJ, 615, 972

\noindent\hangindent=15pt\hangafter=1 Glover, S. C. et al. 2006, ApJ, 640, 553

\noindent\hangindent=15pt\hangafter=1 Grillmair, C. J. et al. 2008, Nature, 456, 767

\noindent\hangindent=15pt\hangafter=1 Guillot, T. 1999,  Science, 286, 7277  

\noindent\hangindent=15pt\hangafter=1 Herczeg et al. 2002, ApJ, 572, 310

\noindent\hangindent=15pt\hangafter=1 Imanishi, K. et al.  2001, ApJ, 557, 747

\noindent\hangindent=15pt\hangafter=1 Ichimaru, S. \& Kitamura,
H. 1999, Phys Plasma, 6, 2649

\end{minipage}\hfill
\begin{minipage}[t]{3.1in}

\vspace{-0.2in}
\noindent\hangindent=15pt\hangafter=1 Ji, H. et al. 1998, PRL, 80, 3256

\noindent\hangindent=15pt\hangafter=1 Ji, H. et al. 2006, Nature, 444, 343

\noindent\hangindent=15pt\hangafter=1 Jonkheid, B. et al. 2004, A\&A, 428, 511

\noindent\hangindent=15pt\hangafter=1 Kastner, J. H. et al. 2002, ApJ, 567, 434

\noindent\hangindent=15pt\hangafter=1 Konacki, M. et al. 2004, ApJ, 609, L37

\noindent\hangindent=15pt\hangafter=1 Kwok, S. et al. 1989, ApJ, 345, L51

\noindent\hangindent=15pt\hangafter=1 Lebedev, S. V. et al. 2002, ApJ, 564, 113

\noindent\hangindent=15pt\hangafter=1 McCarthy, M. C. et al. 2006, ApJ, 652, L141

\noindent\hangindent=15pt\hangafter=1 Murray-Clay, R. et al. 2008, ApJ, accepted

\noindent\hangindent=15pt\hangafter=1 Pascucci, L. et al. 2007, ApJ, 663, 383 

\noindent\hangindent=15pt\hangafter=1 Qi, C. et al. 2006, ApJ, 636, 157

\noindent\hangindent=15pt\hangafter=1 Ren, Y. et al. 2005, PRL, 95, 055003

\noindent\hangindent=15pt\hangafter=1 Rho, J. et al. 2008, ApJ, 673, 271

\noindent\hangindent=15pt\hangafter=1 Robinson, T. et al. 2008, AAS, DPS, 40, 1.04

\noindent\hangindent=15pt\hangafter=1 Salama, F. 1999 in Solid Interstellar Matter: The ISO Revolution, Springer-Verlag

\noindent\hangindent=15pt\hangafter=1 Seager, S. \& Sasselov, D. 2000, ApJ, 537, 916

\noindent\hangindent=15pt\hangafter=1 Tielens, A. 2005, The Physics and Chemistry of the Interstellar Medium,  Cambridge U Press

\noindent\hangindent=15pt\hangafter=1 van Ballegooijen, A. 1994, SpaceSciRev, 68, 299

\noindent\hangindent=15pt\hangafter=1 Weir, S. T. et al. 1996, PRL, 76, 18601863  

\noindent\hangindent=15pt\hangafter=1 Zhu, C. et al. 2005, J Phys A, 71, 052710 

\end{minipage}
}

\end{document}